\documentclass[aps,prb,twocolumn,superscriptaddress,longbibliography]{revtex4-2}

\usepackage[colorlinks=true,citecolor=blue]{hyperref}
\usepackage{graphicx}
\usepackage{amsmath}
\usepackage{amssymb}
\usepackage{tabularx}
\usepackage{gensymb}

\DeclareGraphicsExtensions{.png,.jpg,.eps}
\usepackage{xcolor}

\begin{document}

\author{D. Soriano}
\affiliation{Dipartimento di Ingegneria dell'Informazione, Università di Pisa, 56122 Pisa, Italy}

\title{Domain wall formation and magnon localization in twisted chromium trihalides}

\begin{abstract}
The rise of twistronics has revolutionized the field of condensed matter physics, and more specifically the future applications of two-dimensional materials. At small twist angles, the microscopic world becomes strongly correlated, and unexpected physical phenomena such as superconductivity emerge. For magnetic layers, stacking plays a crucial role in the magnetic exchange coupling between the layers leading to non-trivial spin configurations and flat spin-wave dispersion when twisted. In this work, we give a short overview of the most recent theoretical and experimental works reporting the effect of twist angles on two-dimensional magnets. Besides, we discuss the effect of the twist angle and the local antiferromagnetic interlayer exchange coupling on the formation of antiferromagnetic domains in chromium trihalides. Finally, we  show some preliminary results on the effect of the stacking and the twist angle on the spin-wave dispersion of bilayer CrI$_3$.
\end{abstract}

\date{\today}

\maketitle

\section{Introduction}

The first works on twisted 2D layers date back from the 90s. The observation of long periodicities in graphite by scanning tunneling microscopy (STM) experiments were associated with the Moiré pattern formed by small rotations of the top graphene layer of bulk graphite\cite{RongKuiper1993}. A decade later, other STM experiments showed similar results on epitaxially-grown few-layer graphene on SiC\cite{HassConrad2008}. These observations led to a plethora of theoretical works predicting a dramatic reduction of the Fermi velocity ($v_{\rm F}$) and the formation of flat bands in slightly twisted bilayer graphene\cite{GuineaPeres2006,SantosNeto2007,MorellBarticevic2010,BistritzerMacDonald2011}. The first experiment addressing the electronic properties of twisted bilayer graphene was reported by Li \emph{et al.} in 2010 using scanning tunneling spectroscopy (STS). They observed the shift of Van Hove singularities towards the Fermi level while decreasing the twist angle\cite{LiAndrei2010}. In 2018, Cao \emph{et al.} reported for the first time the observation of intrinsic unconventional superconductivity in tiny "magic angle" twisted bilayer graphene samples using a four-probe transport device\cite{CaoJarillo2018,CaoJarillo2018-2}. The strong correlated behaviour of the flat bands formed at these tiny angles are prone to generate a Mott-like insulating behaviour and other exotic many-body phases. Very recently, these twisted samples were studied with very high precision using scanning magnetometry, leading to the discovery of orbital ferromagnetism in twisted graphene samples\cite{TschirhartYoung2021}.

The observation of superconductivity in magic-angle bilayer graphene coincides with the discovery of long-range magnetic order in exfoliated samples of the layered 
magnets CrI$_3$\cite{HuangJarillo2017} and CrGeTe$_3$\cite{GongZhang2017}. This was a long-sought achievement that strongly improved the 2D materials library. The exfoliation and isolation of monolayers of these magnetic materials have made possible the fabrication of different proof-of-principle van der Waals magnetic devices, ranging from ultrathin magnetic tunnel junctions\cite{GongZhang2017,KleinJarillo2018,SongXu2018} to topological superconductors\cite{KezilebiekeLiljeroth2020} and valley excitons\cite{LyonsTartakovskii2020}. However, one of the most interesting features of these materials is their stacking dependence magnetism, which has been thoroughly studied in the case of the chromium trihalide family of 2D magnets\cite{SivadasXiao2018,SorianoRossier2019,JiangJi2019,JangMyung2019}, together with their electric field tunability\cite{JiangMak2018,JiangShan2018,HuangXu2018}, which opens the door for the next generation of fully-electrical spintronic devices.       

The fact that interlayer exchange interactions are dependent on the stacking configuration of 2D magnetic layers, makes these materials very interesting for the exploration of non-collinear magnetism at very small twist angles, which could serve as the starting point for the use of these materials in quantum information technologies. Also, the spin-wave dispersion in twisted magnetic samples could lead to the localization of magnons at very small twist angles which could be used for information storage. In this short review, I will describe the theory of twisted magnets and discuss the most recent experimental progress in the field. Finally, I will give some prospects about the future of magnetic twistronics.      

\section{Theory}
\begin{figure*}[t!]
\centering
    \includegraphics[width=\textwidth]{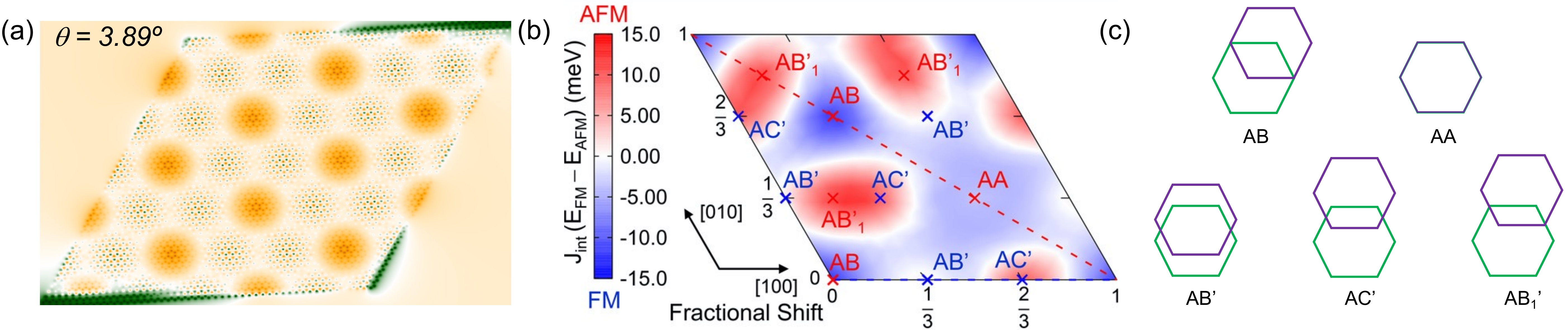}
\caption{(a) Moiré pattern of a $\theta = 3.89$ twisted bilayer hexagonal lattice. Orange, orange/green dotted and white regions belong to AA, AB, and monoclinic (AB', AC', AB'$_1$) stacking configurations respectively. (b) Expected interlayer exchange ($J$) for different stacking configurations in bilayer CrI$_3$ [{\it Reprinted (adapted) with permission from Nano Lett. 2018, 18, 12, 7658–7664. Copyright 2018 American Chemical Society}]. (c) Illustration of the different stacking configurations shown in (b). Green and purple hexagons belong to different layers. }
\label{fig:fig1}
\end{figure*}
The coordinates of a twisted layer can be obtained by a rotation $\theta$ around the $z$-axis, perpendicular to the layer plane:
\begin{equation}
R_z(\theta) = 
\begin{pmatrix}
\cos{\theta} & -\sin{\theta} \\
\sin{\theta} & \cos{\theta}  
\end{pmatrix}
\end{equation}
which transforms the coordinates as
\begin{eqnarray}
x' & = & x\cos{\theta} - y\sin{\theta} \\
y' & = & x\sin{\theta} + y\cos{\theta} 
\end{eqnarray}
In the case of hexagonal homo-bilayers, i.e. bilayers of the same material, commensurate twisted angles are given by the mathematical expression
\begin{equation}
\cos{\theta} = \frac{n^2+4nm+m^2}{2(n^2+nm+m^2)} 
\end{equation}
where $m$ and $n$ are the integer coordinates of the basis vectors $\vec{a}_1 = (1/2,\sqrt(3)/2)a_0$ and $\vec{a}_2 = (-1/2,\sqrt(3)/2)a_0$, and $a_0$ is the lattice parameter. The superperiodicity ($L$) and the lattice vectors of the twisted cell ($\vec{t}_1$ and $\vec{t}_2$) are then determined by the pair of integers $(m,n)$ by
\begin{eqnarray}
L & = & d\sqrt{3(n^2+nm+m^2)} \\
\vec{t}_1 & = & m\vec{a}_1 + n\vec{a}_2 \\
\vec{t}_2 & = & -m\vec{a}_1 + (n+m)\vec{a}_2
\end{eqnarray}
where $d$ is the distance between lattice sites.  

In Fig. \ref{fig:fig1}(a), we show a twisted hexagonal bilayer structure with $\theta = 3.89\degree$, $(m,n) = (9,8)$. For small angles, the Moiré patterns show regions with a clear stacking, namely, AA (in orange), AB (in dotted orange/green). These regions have shown to be very important in understanding the electronic properties of twisted graphene\cite{MorellBarticevic2010,BistritzerMacDonald2011,SanJosePrada2013} and transition metal dichalcogenides (TMD)\cite{SorianoLado2020,SorianoLado2021}, since the states forming the strongly correlated flat bands in these materials are mostly localized in AA or AB regions. In 2D magnetic materials, these ordered stacking configurations are connected to a given interlayer magnetic ordering, namely, ferromagnetic (FM) or antiferromagnetic (AFM). Fig. \ref{fig:fig1}(b) shows the expected interlayer magnetic ordering for the different stacking configurations in bilayer CrX$_3$ shown in Fig. \ref{fig:fig1}(c). The Moiré pattern for $\theta = 3.89\degree$ shows that most of the stacking is AA or AB leading to an overall FM interlayer magnetism. However, the regions connecting the AA domains (in white) show monoclinic stacking configurations such as AB', AC', and AB'$_1$, where an AFM interlayer magnetism is expected. However, the formation of these AFM domains is governed by certain parameters like the single-ion anisotropy or the AFM interlayer exchange as described in the next section.

\subsection{Domain wall formation in twisted chromium trihalides}
The formation of AFM domains in a twisted FM structure is not easy to attain and depends on the difference between the energy gained for having an AFM ($E_{AFM}$) ordering in a monoclinic region, and the one needed to form a domain wall ($E_{DW}$):\cite{AkramBotana2021}
\begin{eqnarray}
& E & = E_{AFM} - E_{DW} \approx 2f_{AFM}\left(\frac{a_M}{a}\right)^2J_{\perp} - \nonumber \\
& &  \pi\left(\frac{a_M}{a}\right)\sqrt{(J_{\parallel}+\frac{\lambda}{2})(K+\lambda+0.72J_\perp)}
\label{eqn:formation}
\end{eqnarray}
where $f_{AFM}$ is the fraction of atoms with an AFM interaction, $a_M$ and $a$ are the Moiré and monolayer lattice parameters, $J_\perp$ and $J_\parallel$ are the interlayer and intralayer exchange, $\lambda$ is the anisotropic exchange coupling term (mediated by the ligands), and $K$ the single-ion anisotropy. The domain wall width ($\delta$) is given by:
\begin{equation}
\delta \approx \pi\sqrt{\frac{J_\parallel+\frac{\lambda}{2}}{K+\lambda+0.72J_\perp}}a
\label{eqn:delta}
\end{equation}
where it is clearly seen the competing effect of anisotropy and intralayer exchange on the domain wall size. 

\begin{table}[]
    \caption{In-plane and inter-layer exchange coupling ($J$) values of bilayer chromium trihalides obtained from density functional theory calculations. Single-ion anisotropy ($K$) and magnetic anisotropic exchange ($\lambda$) calculated by Akram \emph{et al.}\cite{AkramBotana2021}}
    \centering
    \begin{tabularx}{\columnwidth}{X X X X X}
        \hline
         & $J_\perp$ (meV) & $J_\parallel$ (meV) & $\lambda$ (meV) & $K$ (meV)  \\ 
         \hline
         CrI$_3$ & -0.088 & 2.2 & 0.14 & 0.03 \\
         CrBr$_3$ & -0.038 & 1.4 & 0.04 & 0.03 \\
         CrCl$_3$ & -0.001 & 0.9 & 0.003 & 0.006 \\
         \hline
    \end{tabularx}
    \label{tab:Jcoupling}
\end{table}

\begin{figure}[t!]
\centering
    \includegraphics[width=\columnwidth]{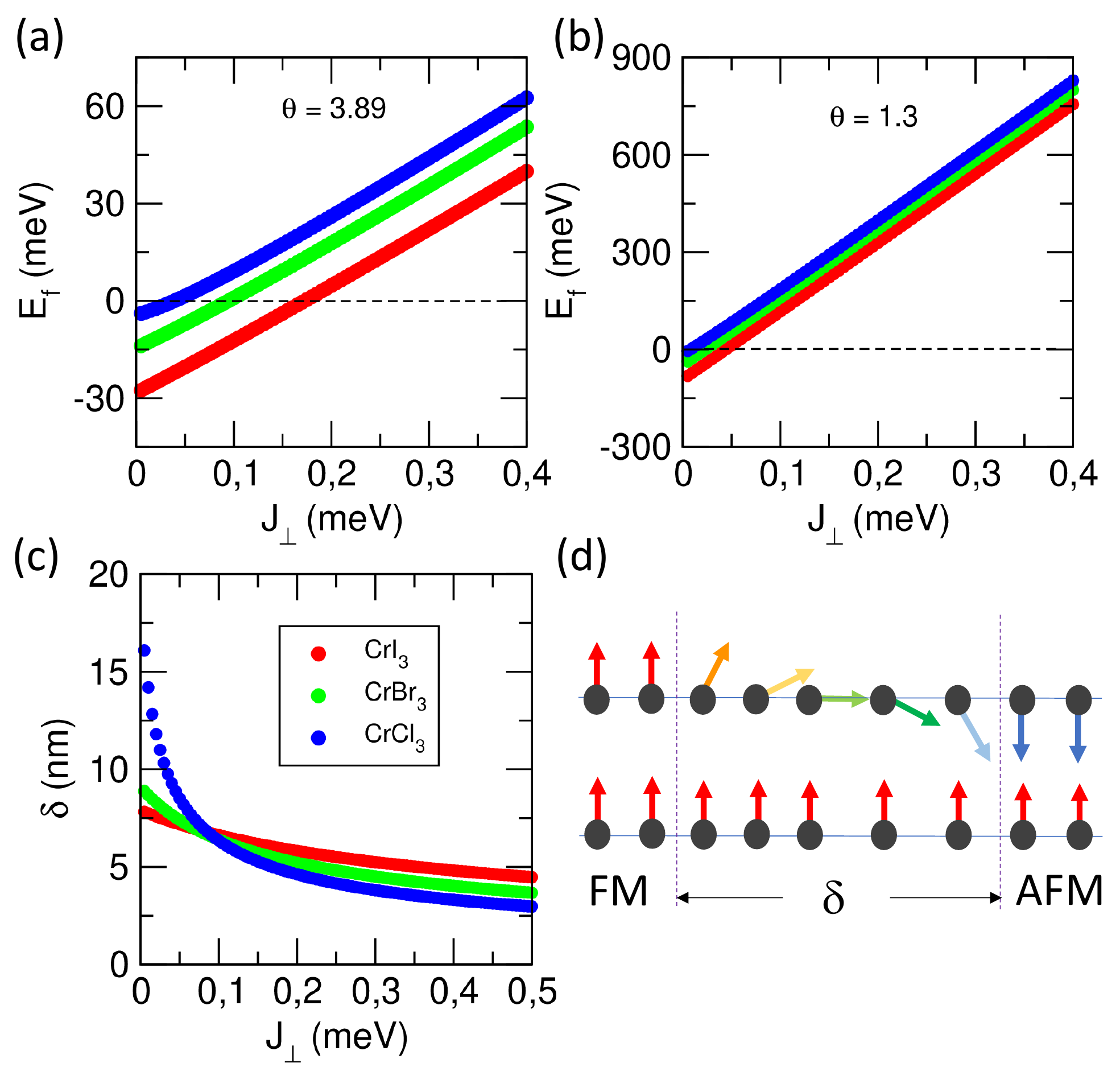}
\caption{Formation energy of AFM domains with increasing AFM interlayer exchange ($J_\perp$) of bilayer chromium trihalides for twist angles (a) $\theta = 3.89\degree$ and (b) $\theta = 1.3\degree$. (c) Domain wall thickness for the three different chromium trihalides with increasing $J_\perp$. (d) Schematics of domain wall formation between FM and AFM domains.}
\label{fig:fig2}
\end{figure}

In Fig. \ref{fig:fig2}, we show the expected formation energy and domain wall thickness for different values of the AFM interlayer exchange coupling ($J_\perp$) and for two different twist angles, namely $\theta = 3.89\degree$ and $\theta = 1.3\degree$ $(n,m) = (26,25)$, using the intralayer exchange coupling parameters ($J_\parallel$) and anisotropies ($K$ and $\lambda$) reported on Table \ref{tab:Jcoupling}, $f_{AFM} = 0.5$, and $D = 0$. The results show that the formation of AFM domains for the case $\theta = 3.89\degree$ is forbidden for the typical $J_\perp$ values reported on Table \ref{tab:Jcoupling} due to the small $a_M/a$ ratio, while it is mostly favoured in the case $\theta = 1.3\degree$ for CrI$_3$ and CrBr$_3$. For CrCl$_3$, the small interlayer exchange coupling ($J_\perp$) makes difficult the formation of AFM domains. Also, by looking at Fig. \ref{fig:fig2}(c), the domain wall thickness in the case of CrCl$_3$ increases dramatically close to $J_\perp = 0$. Considering that the typical $|J_\perp|$ for this material is $\sim  0.001$ meV, the size of the domain wall becomes comparable to the Moiré lattice vector ($a_M \sim 30$ nm) making very difficult the formation of a domain wall. For CrI$_3$ and CrBr$_3$, $\delta$ ranges between $5$ to $10$ nm, around three times smaller than $a_M$, which may favor the formation of a domain wall in twisted bilayer samples. Nonetheless, it is important to note how these results are sensitive to the local variations of the AFM interlayer exchange ($J_\perp$) originated from the different monoclinic stacking regions in twisted samples.  In fact, Fig. \ref{fig:fig1}(b) shows that the typical monoclinic stacking (AB') is not the one with the higher $J_\perp$. For instance, AC' and AB'$_1$ stackings show much stronger $J_\perp$ which can eventually decrease the domain wall size locally, increasing the probability of observing AFM domain walls in twisted CrI$_3$ and CrBr$_3$, and even CrCl$_3$.

On the other hand, the formation of non-collinear phases and skyrmions in chromium trihalides have been recently predicted using stochastic Landau-Lifshitz-Gilbert simulations and DFT calculations \cite{GhaderStroppa2021,AkramBotana2021,XiaoTong2021}. In presence of spin-orbit coupling, layered structures with broken inversion symmetry can give rise to an antisymmetric exchange anisotropy, also known as Dzyaloshinskii-Moriya interaction (DMI). Recent experiments have observed the formation of spin spirals in the surface of Fe$_3$GeTe$_2$ induced by the strong DMI\cite{MeijerGuimaraes2020}. These spin spirals show periodicities longer than the crystal (or Moiré) lattice vector which opens new possibilities for the exploration of non-trivial spin phenomena in the situations where domain walls exceed the Moiré lattice vector. 

\subsection{Magnon dispersion in twisted chromium trihalides}
The excitation of magnons in 2D magnets is one of the most exciting properties of these materials, which may lead to the development of low-power spintronic devices where spins are transmitted through magnons (or spin-waves) instead of electrons or holes\cite{KleinJarillo2018,GhazaryanMisra2018,LiuWees2020}. Twisting can become a powerful tool to tune the properties of magnons in 2D magnets.\cite{HejaziBalents2020} 

\begin{figure}[t!]
\centering
    \includegraphics[width=\columnwidth]{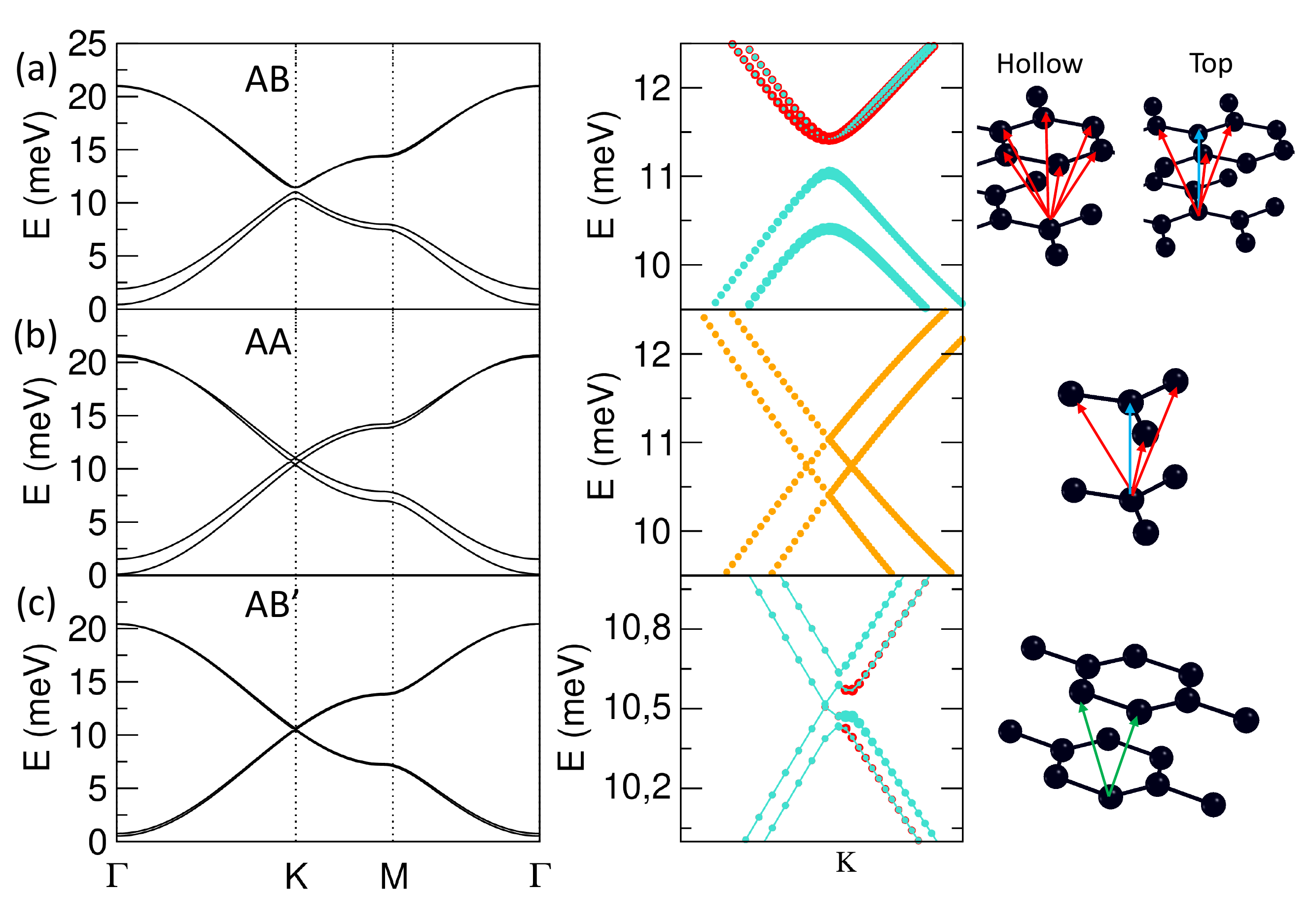}
\caption{Spin-wave dispersion of bilayer CrI$_3$ for (a) AB, (b) AA, and (c) AB' stacking configurations. On the right panels, the region around the Dirac point is magnified together with the projection on the two sublattices (green and red circles). Orange circles in the AA case denotes that both sublattices are equivalent. The detail of the atomic coordination for each stacking is also depicted. Red arrows indicate a FM coupling $J_\perp^{(2)} \sim 0.09$ meV from the hollow site to the six nearest neighbors in the opposite lattice. For the top site, the blue arrow indicates the AFM nearest-neighbor coupling between lattices $J_\perp^{(1)} \sim -0.21$ meV, together with 3 next nearest-neighbour coupling of type $J_\perp^{(2)}$. For the AA case, the coupling between layers is the same as the top configuration in the AB stacking. Finally, the AB' stacking shows an interpolated AFM $J_\perp^{(3)}$ in green arrows where each atom is coordinated to two atoms int eh opposite layer.}
\label{fig:fig3}
\end{figure}

To obtain the magnon dispersion, it is important to describe properly the interaction between spins in the crystal. For chromium trihalides, the most simple spin Hamiltonian for bilayer CrX$_3$ is given by:
\begin{eqnarray}
\label{eqn:ham}
\mathcal{H}_{spin} & = & \sum_i K\left(S_i^z\right)^2 + \frac{1}{2}\sum_{i,j}J_{\parallel,ij}\mathbf{S_i}\mathbf{S_j} + \nonumber \\
& & \frac{\lambda}{2}\sum_{\langle i,j\rangle} S^z_i S^z_j +
\frac{1}{2}\sum_{i,j}J_{\perp,ij}\mathbf{S_i}\mathbf{S_j}   
\end{eqnarray}
where $K$ is the single-ion anisotropy, $S_i^z$ is the $z$-component of the spin at site $i$, $J_\parallel$ and $j_\perp$ are the intra-layer and inter-layer exchange coupling, $\mathbf{S_i}$ is the spin vector at each lattice site, and $\lambda$ is the magnetic anisotropic exchange between nearest Cr atoms. Given that second- and third-neighbor intra-layer exchange have small effects on the magnon dispersion close to the Dirac point, we only consider nearest-neighbor exchange interactions.

The spin-wave Hamiltonian can be obtained by transforming the spin Hamiltonian in Eq.\ref{eqn:ham} using Holstein-Primakoff:
\begin{eqnarray}
\mathcal{H}_{SW} & = & S\sum_i \left[(K+3(J_\parallel+\lambda)+\tilde{J}_{\perp,i})\right]b_i^\dagger b_i - \nonumber \\
& & J_\parallel S\sum_\nu\sum_{\left<i,j\right>}b_{\nu i}^\dagger b_{\nu j} - J_\perp S\sum_\nu\sum_{i,j}b_{\nu i}^\dagger b_{\bar{\nu}j}     
\end{eqnarray}
where $\tilde{J}_{\perp,i} = \sum_{j \neq i} J_{\perp,i,j}$ depends on the coordination of each atom with the neighboring atoms in the opposite layer (see Fig.\ref{fig:fig3}), $\nu$ is the layer index, and $b_1$, $b_2$, $b_1^\dagger$, $b_2^\dagger$ are the boson creation and annihilation operators for sites in layers 1 and 2. For the interlayer coupling, we use the values reported by Sivadas et al.\cite{SivadasXiao2018}, where the interlayer coupling depends on the distance to the atoms on the opposite layer. For top configuration, the interlayer coupling is AFM $J_\perp^{(1)} = -0.21$ meV, while for the hollow sites the coupling is FM $J_\perp^{(2)} = 0.09$ meV. For distances between top and hollow, such as those observed in the AB' configuration, we use an interpolated value $J_\perp = (1-d/l)J_\perp^{(1)}+(d/l) J_\perp^{(2)}$, where $d$ is the in-plane distance between inter-layer neighbors, and $l$ is the inter-atomic distance.  

In Fig.\ref{fig:fig3} (left panel), we show the magnon dispersion for three different stacking configurations AA, AB and AB'. The dispersions resemble the typical electronic structure of bilayer graphene, with parabolic bands at the Dirac points. In the right panel of Fig.\ref{fig:fig3}, the region around the Dirac point is magnified to show the dispersion of the magnon bands near the K-point. Interestingly, we find that a small gap opens at K of the order of $\sim 0.3$ meV for AB stacking. Similar gaps have been reported in previous calculations for this stacking configuration\cite{OrtmannsBlanter2021} and bulk crystals\cite{KeKatsnelson2021}. The projection of the bands shows a change from the low-energy (acoustic) branch and the high-energy (optical) branch at K-point. While in the acoustic branch the states are localized in the top sites (cyan circles), in the optical branch the states are localized in the hollow sites (red circles). For the AA stacking, the bands at the K-point show a gapless linear dispersion with two Dirac cones shifted $\sim 0.5$ meV. In this case, all the states are equally weighted in all lattice sites (orange circles). Finally, for the AB' stacking, we get two different Dirac cones similar to the AA case. However, one of them at K-$\delta$ is gapless while the other one at K+$\delta$ shows a very small gap $\sim 0.1$ meV with similar behavior to the AB stacking. These results show that the spin-wave dispersion close to the Dirac point in bilayer CrX$_3$ may be used as a footprint to identify the different stacking configurations. 

\begin{figure}[t!]
\centering
    \includegraphics[width=\columnwidth]{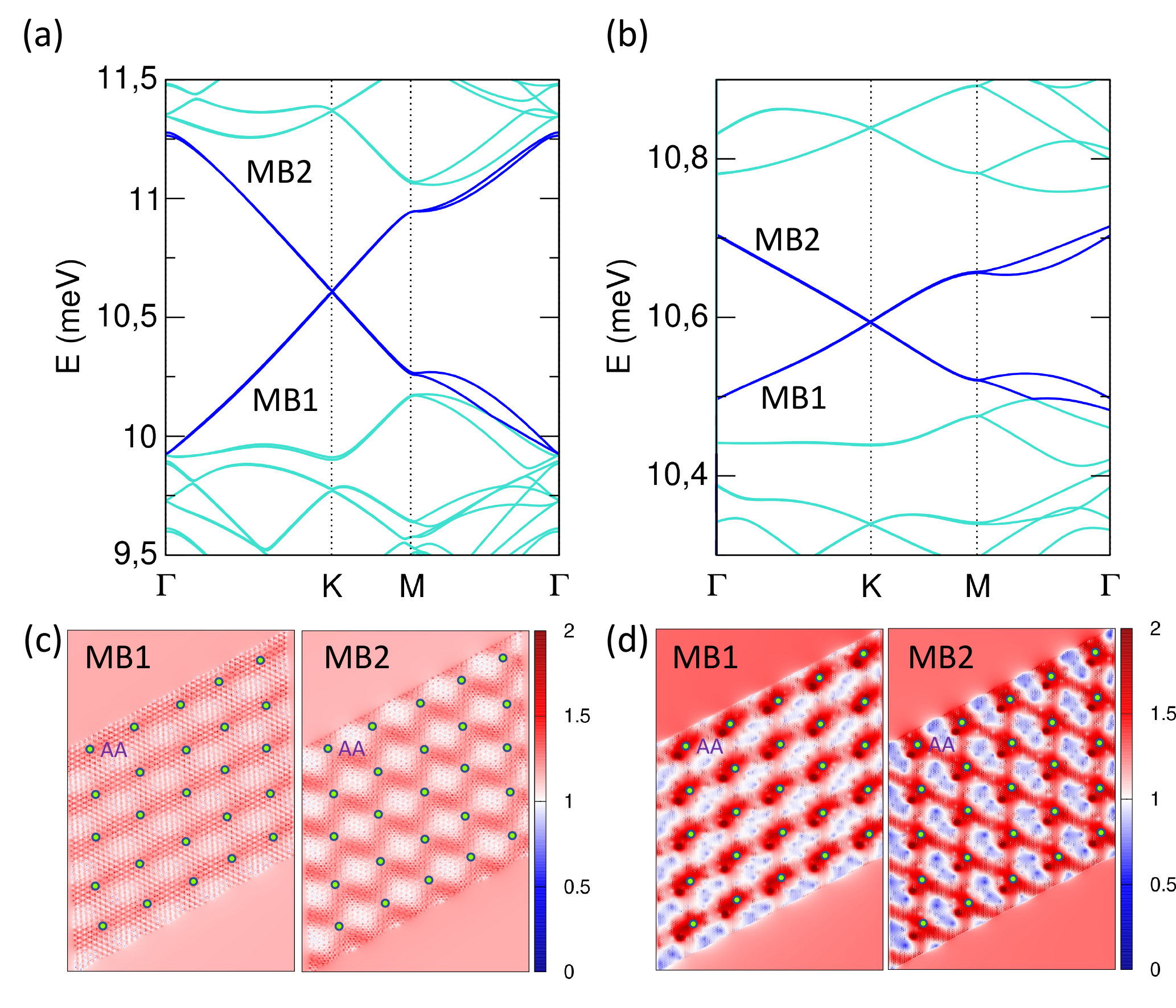}
\caption{Spin-wave dispersion for (a) $\theta = 3.89\degree$ and (b) $\theta = 1.3\degree$. (c) Projections of the two magnon bands close to the Dirac point (MB1 and MB2) for $\theta = 3.89\degree$. (d) Same as (c) for $\theta = 1.3\degree$. The projections are shown in a 5x5 twisted supercell. Green dots mark the position of the AA stacking regions. }
\label{fig:fig4}
\end{figure}

The effect of twist angles in the spin-wave dispersion of 2D magnets is shown in Fig.\ref{fig:fig4}. Similar to the case of the electronic structure of twisted 2D materials, the dispersion flattens when reducing the twist angle. The bandwidth changes from $1.3$ meV for $\theta = 3.89\degree$ to $0.2$ meV for $\theta = 1.3\degree$ for the magnon band below the Dirac point (MB1) (see Fig. \ref{fig:fig4}(a,b)). In Fig.\ref{fig:fig4}(c,d), we show the real space projection of the two magnon bands below (MB1) and above (MB2) the Dirac point. In both cases, the magnons are localized in the AA sites (green dots), but with higher intensity in the $\theta = 1.3\degree$ case. Also, depending on the band, the magnons extend along different monoclinic directions connecting the AA sites. These 1D pathways connect FM and AFM domains along domain walls. Since domain walls could be used to store information, these magnon pathways may facilitate the transport of information through twisted 2D magnetic materials.

\section{Experiments}
\begin{figure}[t!]
\centering
    \includegraphics[width=\columnwidth]{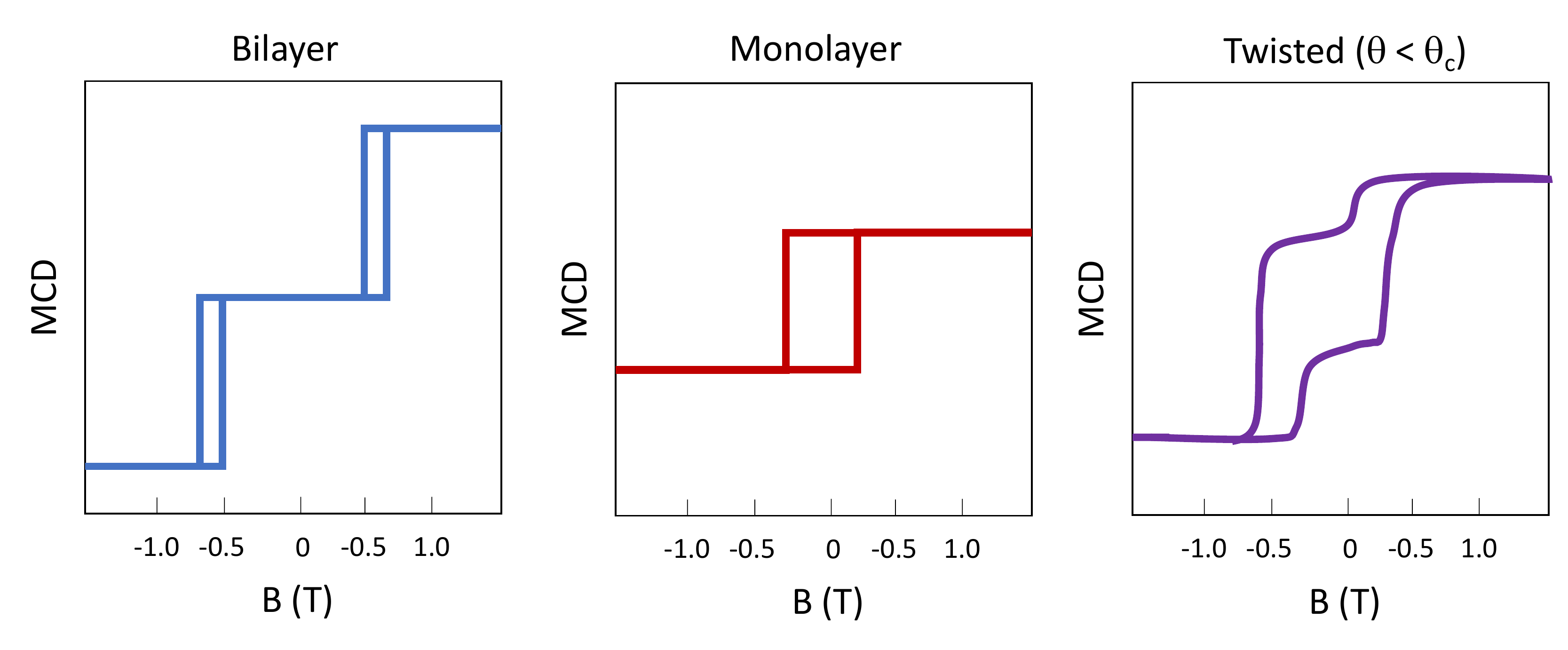}
\caption{Illustrative picture of the RMCD signal of twisted CrI$_3$ for $\theta < \theta_c$(c). It resembles a combination of the monolayer (b) and bilayer (c) CrI$_3$ RMCD loop signals.  }
\label{fig:fig5}
\end{figure}

The experimental observation of Moiré superlattices on twisted 2D magnets was first reported by Song \emph{et al.}\cite{SongXu2021} using single-spin quantum NV center magnetometry\cite{ThielMaletinsky2019}, which allows for high spatial resolution ($\sim 50$ nm) and sufficient magnetic sensitivity. The use of quantum magnetometry has made possible the observation of the coexistence of FM and AFM domains in CrI$_3$ $\theta = 0.2\degree$ twisted samples. On the other hand, experiments based on magneto-optical probes such as polar reflective magnetic circular dichroism (RMCD) have been also used to elucidate the magnetic states of 2D magnetic materials, however its resolution is limited by the laser spot size ($\sim 1 \mu$m). A close look into the evolution of the hysteresis loops obtained by RMCD with the twist angle shows that the coexistence of FM and AFM domains disappears for a critical angle $\theta_c = 3\degree$\cite{XuShan2022}. For twist angles below the critical one, the RMCD signal shows a distorted hysteresis loop resembling a combination of the monolayer and bilayer signals as shown schematically in Fig. \ref{fig:fig5}.

A very interesting property of 2D magnets is the possibility to tune the interlayer exchange coupling with an external electric field. The weak inter-layer exchange interaction in the AFM gound state ($J_\perp < 0.1$ meV) makes easy to switch between AFM and FM inter-layer coupling with. In this regard, the fraction of atoms with FM inter-layer interaction in twisted magnets can be easily tuned with a gate voltage ($-40$ to $40$ V) reducing the AFM inter-layer exchange coupling and allowing a phase transition from the non-collinear to the collinear FM state\cite{XuShan2022}.

Although there have been great advances in the experimental observation of magnetic domains in twisted 2D magnetic materials, there are still some issues that needs to be solved to get a better insight on the formation of these complex magnetic structures. One of the main drawbacks regarding the experimental observation of Moiré magnets in chromium trihalides is related to the possibility of lattice reconstruction at small twist angles and the emergence of local structural deformations during the fabrication process\cite{SongXu2021,XieZhao2022,WangXiao2020}. More specifically, at small twist angles, van der Waals interactions can induce interface structural reconstructions favoring interlayer commensurability as already reported for twisted graphene\cite{YooKim2019}. These effects may lead to the formation of disordered magnetic structures and the lack of periodic magnetic domain patterns.

\section{Summary and Prospect}
In this short review, we tried to give an overview of the recent theoretical and experimental advances regarding twisted chromium trihalides. We have described how magnetic anisotropy and interlayer AFM exchange play a very important role in the coexistence of FM and AFM domains. Additionally, the spin-wave dispersion of bilayer CrI$_3$ for different stacking configurations and twisted samples has been calculated using a detailed inter-layer exchange interaction parameter. Interestingly, we found that the spin-wave dispersion at K depends strongly on the stacking, and become localized along the monoclinic stacking regions for twisted samples. Finally, the few, but very recent, experiments on twisted chromium trihalides have been discussed. 

The twistronics era has reached all kinds of two-dimensional materials, including 2D magnetic materials. As described in this short review, the possibility to induce the coexistence of FM and AFM domains by small twist angles in chromium trihalides allows for the exploration of magnonics applications at the 2D limit. Probably, one of the most exciting applications could be the possibility to use domain walls and spin-waves to store and carry information\cite{JiahaoLuqiao2019}. Usually, this is limited by the short propagation of the spin waves. However, recent theoretical and experimental works have demonstrated the presence of topological magnons at the edges of CrI$_3$ films with longer coherence lengths\cite{ChenDai2018,CostaRossier2020,ChenDai2021}. Also, the formation of flat spin-wave bands at small twist angles, potentially connected with localized magnons, together with the high electrical tunability of the non-collinear magnetic structures, may open new paths for the application of these materials in novel quantum information technologies\cite{YuanPeng2021,BarmanWinklhofer2021}. 

Nevertheless, to reach the goal of 2D magnonics, some issues regarding fabrication and potential interface reconstruction should be overcome to obtain well-ordered magnetic domains in twisted samples. At the same time, an exhaustive search of new 2D magnets with high Curie temperatures and able to be handled at room temperature is becoming a key step to make these materials technologically useful in future applications.

\textbf{Acknowledgments:} D.S. acknowledges fruitful discussions with Mikhail Titov and Efrén Navarro.

\bibliography{biblio}{}
\end{document}